%% file: metrics-survey.tex
\documentclass[prodmode]{acmsmall} 

\usepackage{import}
\usepackage{amsmath, amssymb, mathrsfs, bm, latexsym,graphics,color, graphicx,url}
\usepackage{subfigure}
\usepackage{graphicx}
\usepackage{epstopdf}
\usepackage{longtable}
\usepackage{rotating}

\newcommand{\A}{{\ensuremath{\cal A}}}
\newcommand{\op}{{\bf o}}

\newcommand{\B}{{\cal B}}
\newcommand{\R}{{\mathbb R}}

\newcommand{\D}{{\ensuremath{\cal D}}}

\newcommand{\ignore}[1]{}

\begin{document}

\markboth{M. Pendleton et al.}{A Survey on Security Metrics}

\title{A Survey on Security Metrics\footnote{Author's addresses: M. Pendleton {and} R. Lebron-Garcia {and} S. Xu, Department of Computer Science,
The University of Texas at San Antonio. Correspondence: Shouhuai Xu ({\tt shxu@cs.utsa.edu})
}
}
\author{MARCUS PENDLETON
\affil{The University of Texas at San Antonio}
RICHARD GARCIA-LEBRON
\affil{The University of Texas at San Antonio}
SHOUHUAI XU
\affil{The University of Texas at San Antonio}
}

\begin{abstract}
\input{abstract.tex}
\end{abstract}

%
%

%

%
%






\maketitle

\section{Introduction}

Security metrics is one of the most important open problems in security research.
It has been recognized on the  Hard Problem List of the United States INFOSEC Research Council
(both 1999 and 2005 editions) \cite{IRC-hardproblemlist},
has been reiterated in 2011 by the United States National Science and Technology Council \cite{NITRD},
and most recently has been listed as one of the five hard problems in Science of Security (August 2015) \cite{NSAHardProblemList}.

The security metrics problem certainly has received a lot of attention, including government and industry bodies \cite{NIST800-55Rev1,IATAC-Report-2009,SANS,CIS}.
For example, the United States National Institute of Standards and Technology
proposed three categories of security metrics---{\em implementation}, {\em effectiveness}, and {\em impact}  \cite{NIST800-55Rev1};
the Center for Internet Security defined 28 security metrics in another three categories---{\em management}, {\em operational}, and {\em technical} \cite{CIS}.
However, these efforts are almost exclusively geared towards cyber defense administrations and operations.
They neither discuss how the security metrics may be used as parameters in security modeling (i.e., theoretical use of security metrics),
nor discuss what the gaps are between the state-of-the-art and the ultimate goals and how these gaps may be bridged.
This motivates us to survey the knowledge in the field, while hoping
to shed some light on the difficulties of the problem and the directions for future research.
To the best of our knowledge, this is the first survey of security metrics,
despite that there have been some efforts with a much narrower focus
(e.g., \cite{Landwehr:1994:TCP:185403.185412,Chandola:2009:ADS:1541880.1541882,Milenkoski:2015:ECI:2808687.2808691,Roundy:2013:BOP:2522968.2522972,DBLP:conf/sp/Ugarte-PedreroB15}).

The paper is organized as follows.
Section \ref{sec:methodology} discusses the scope and methodology of the survey.
Section \ref{sec:knowing-your-vulnerabilities} describes security metrics for measuring system vulnerabilities.
Section \ref{sec:knowing-your-defense} reviews security metrics for measuring defenses.
Section \ref{sec:knowing-your-enemy} presents security metrics for measuring threats.
Section \ref{sec:knowing-the-situation} describes security metrics for measuring situations.
Section \ref{sec:discussion} discusses the gaps between the state-of-the-art and the security metrics that are desirable.
Section \ref{sec:conclusion} concludes the paper.

\section{Scope and Methodology}
\label{sec:methodology}

\subsection{Terminology}

The term {\em security metrics} has a range of meanings, with no widely accepted definition \cite{Jansen09directionsin}.
It is however intuitive that security metrics reflect some security attributes quantitatively.

Throughout the paper, the term
{\em systems} is used in a broad sense, and is used in contrast to the term {\em building-blocks}, used to describe concepts  such as cryptographic primitives.
The discussion in the present paper applies to two kinds of systems:
(i) {\em enterprise systems}, which include networked systems of multiple computers/devices (e.g., company networks), clouds, and even the entire cyberspace, and
(ii) {\em computer systems}, which represent individual computers/devices.
This distinction is important because
an enterprise system consists of many computers/devices, and measuring security of an enterprise system naturally requires to measure security of the individual computers.

The term {\em attacking computer} represents a computer or IP address from which cyber attacks are launched against others,
while noting that the attacking computer itself may be a compromised one (i.e., not owned by a human attacker).
The term {\em incident} represents a successful attack (e.g., malware infection or data breach).

For applications of security metrics, we will focus on two uses.
The {\em theoretical use} is to incorporate security metrics as parameters into some security models
that may be built to understand security from a more holistic perspective.
There have been some initial studies in pursuing such models, such as \cite{SandersQUEST2011,XuCybersecurityDynamicsHotSoS2014},
which often aim to characterize the evolution of the global security state.
The {\em practical use} is to guide daily security practice, such as comparing the security of two systems
and comparing the security of one system during two different periods of time (e.g., last year vs. present year).

\subsection{Scope}

We have to limit the scope of the literature that is surveyed in the present paper.
This is because {\em every} security paper that improves upon a previous result---be it a better defense or more powerful attack---would be considered relevant in terms of security metrics.
However, most security publications did {\em not} address the security metrics perspective,
perhaps because it is sufficient to show, for example, a newly proposed defense can defeat an attack that
could not be defeated by previous defenses.
This suggests us to survey the literature that made a reasonable effort at defining security metrics.
This selection criterion is certainly subjective, but we hope the readers find the resulting survey and discussion
informative. It is worth mentioning that our focus is on security metrics, rather than the specific approaches for analyzing them.
We treat the analysis approaches as an orthogonal issue because a security metric may be analyzed via multiple approaches.

Even within the scope discussed above, we still need to narrow down our focus.
This is because security, and security metrics thereof, can be discussed at multiple levels of abstractions, including systems and building-blocks as mentioned above.
For building-blocks, great success has been achieved in measuring the {\em concrete security} of cryptographic primitives \cite{DBLP:conf/focs/BellareDJR97},
while other notable results include metrics for measuring privacy \cite{DBLP:conf/icalp/Dwork06,Shokri:2011:QLP:2006077.2006769}, information flow \cite{ClarksonOakland2014},
side-channel leakage \cite{DBLP:journals/iacr/SchneiderM15}, and hardware security \cite{6860363}.
On the other hand, our understanding of security metrics for measuring security of {\em systems} lags far behind, as the present paper shows.
One thing that is worth clarifying is that the exposure of cryptographic keys, due to the use of weak randomness in the key generation algorithm or Heartbleed-like attacks,
is treated as a systems security problem. This is plausible because the formal framework for analyzing cryptographic security
assumes that the cryptographic keys are not exposed.

The aforementioned discrepancy between the metrics for systems security and the metrics for building-blocks security
is unacceptable because the former
is often needed and used in the process of business decision-making.
This suggests us to focus on systemizing the underdeveloped field of systems security metrics.
The importance of this underdeveloped field can be seen by efforts that have been made by
government and industrial bodies \cite{NIST800-55Rev1,IATAC-Report-2009,SANS,CIS}.
This prompts us to consider both these metrics and those that appeared in academic venues.


\subsection{Survey methodology}

Our survey methodology is centered on the perspective of cyber attack-defense interactions,
which applies to both {\em enterprise systems} and {\em computer systems} mentioned above.

\begin{figure}[!htbp]
\centering
\includegraphics[width=.8\textwidth]{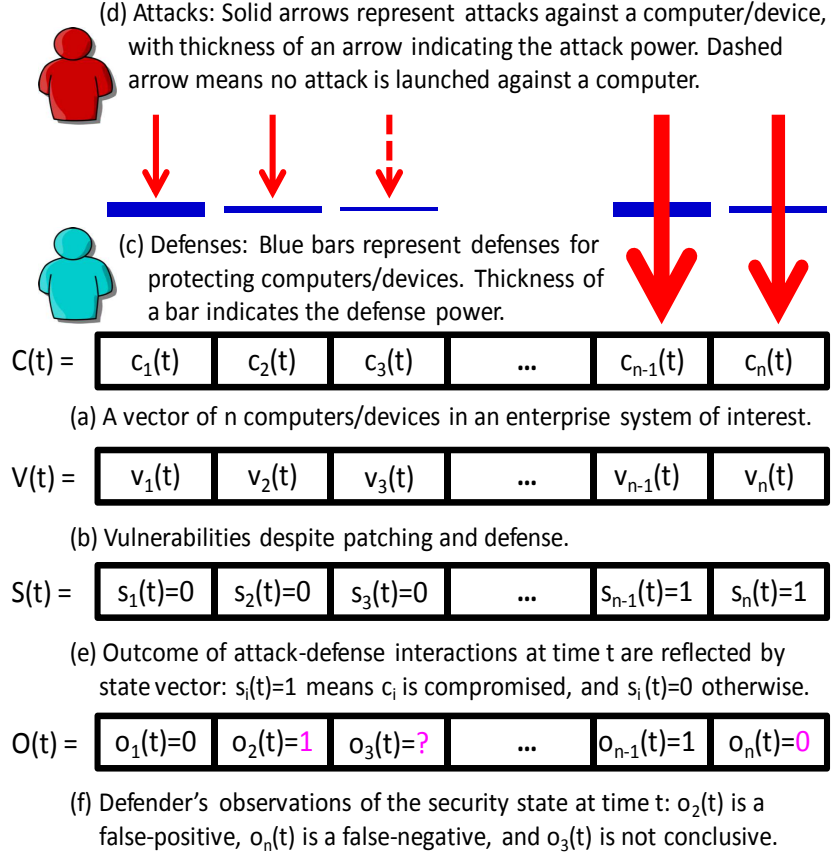}
\caption{An abstract representation of a snapshot of an enterprise system at time $t$, where the defenses (i.e., blue bars)
accommodate both the defenses that are installed on the individual computers
(e.g., anti-malware tools) and the defenses that are employed at the perimeter of the enterprise system (e.g., firewalls).
Some attacks penetrate through the defenses (e.g., attacks against computer $c_n(t)$), while others fail (e.g., attacks against computer $c_1(t)$).
Computer $c_i(t)$ may have a vector $v_i(t)$ of vulnerabilities, some of which may not be known to the defender (i.e., zero-day).
The outcome of the attack-defense interaction at time $t$ is reflected by a security state vector
$S(t)=\{s_1(t),\ldots,s_n(t)\}$, where $s_i(t)=0$ means computer $c_i(t)$ is secure at time $t$ and  $s_i(t)=1$ means computer $c_i(t)$ is compromised at time $t$.
However, the defender's observation of the security state vector $S(t)$, denoted by
$O(t)=\{o_1(t),\ldots,o_n(t)\}$, may not be perfect because of false-positives, false-negatives, or non-decisions.}
\label{fig:taxonomy}
\end{figure}

Figure \ref{fig:taxonomy} illustrates a snapshot of an {\em enterprise system}.
At time $t$, the enterprise system consists of $n$ computers (or devices, virtual machines in the case of cloud), denoted by the vector $C(t)=\{c_1(t),\ldots,c_n(t)\}$,
where $n$ could vary with time $t$ (i.e., $n$ could be a function of time $t$).
Each computer, $c_i(t)$, may have a vector $v_i(t)$ of vulnerabilities,
including the computer user's vulnerability to social-engineering attacks,
the vulnerability caused by the use of weak passwords, and
the software vulnerabilities that may include some zero-day and/or some unpatched ones.
Attacks are represented by red arrows, and
defenses are represented by blue bars.
Defenses accommodate both the defenses that are installed on the individual computers (e.g., anti-malware tools) and the defenses that are employed at
the perimeter of the enterprise system (e.g., firewalls).
The thickness of red arrows and blue bars reflect the attack and defense power, respectively.
Some attacks penetrate through the defenses (e.g., attacks against computer $c_n(t)$), while others fail (e.g., attacks against computer $c_1(t)$).
The outcome of the attack-defense interaction at time $t$ is reflected by a security state vector
$S(t)=\{s_1(t),\ldots,s_n(t)\}$, where $s_i(t)=0$ means computer $c_i(t)$ is secure at time $t$ and  $s_i(t)=1$ means computer $c_i(t)$ is compromised at time $t$.
However, the defender's observation of the security state vector $S(t)$, denoted by
$O(t)=\{o_1(t),\ldots,o_n(t)\}$, may not be perfect because of false-positive, false-negative, or noise.

\begin{figure}[!htbp]
\centering
\includegraphics[width=.8\textwidth]{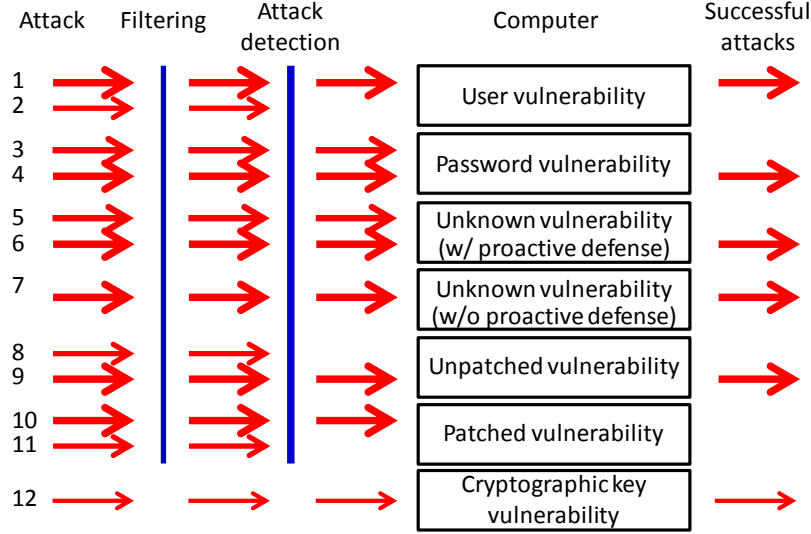}
\caption{A snapshot of attacks against a computer (or device), say $c_i(t)$, in the enterprise system at time $t$,
where blue bars also represent defenses and red arrows represent attacks (their thickness reflect their defense/attack power).
If $c_i(t)$ was compromised at time $t_1<t$ and is not cleaned up at time $t$, or
if $c_i(t)$ is compromised at time $t$, then $s_i(t)=1$. If the defender correctly observes the state of $c_i(t)$ at time $t$, the observation is $o_i(t)=1$.
For cryptographic key vulnerabilities (e.g., Heartbleed-like vulnerabilities that are not patched), there is essentially no defense that can block the attack.
\label{fig:taxonomy-individual-view}}
\end{figure}

Figure \ref{fig:taxonomy-individual-view} illustrates a snapshot of a {\em computer system} $c_i(t)$ at time $t$,
where we also use blue bars to represent defenses, use red arrows to represent attacks, and use their thickness to reflect their defense/attack power.
At a high level, $c_i(t)$ may have a range of vulnerabilities, which correspond to $v_i(t)$ in Figure \ref{fig:taxonomy}.
The vulnerabilities include the computer user's vulnerability (or susceptibility) to social-engineering attacks,
the vulnerability caused by the use of weak passwords, and software vulnerabilities.
The defense may include
(i) the use of some {\em filtering} mechanisms that are deployed at the enterprise system perimeter
to block (for example) traffic from malicious or blacklisted IP addresses,
(ii) the use of some {\em attack detection} mechanisms to detect and block attacks before they reach computer $c_i(t)$, and
(iii) the use of some proactive defense mechanisms (e.g., address space randomization) to try to prevent the exploitation of some vulnerabilities.
Suppose the attacker has a vector of 12 attacks.
Attack 1 successfully compromises $c_i(t)$ because the user is lured into (e.g.) clicking a malicious URL.
Attack 4 successfully compromises $c_i(t)$ because the password in question is correctly guessed.
Attacks 6 and 7 successfully compromise $c_i(t)$ because they exploit a zero-day vulnerability, despite the possible employment of proactive defense mechanisms on $c_i(t)$.
Attack 9 successfully compromises $c_i(t)$ because the vulnerability is unpatched and the attack is neither filtered nor detected.
Attack 12 successfully compromises $c_i(t)$ because the cryptographic key in question is exposed by (e.g.,) Heartbleed-like attacks against which no defense is employed
(i.e., the lack of blue bars).
All of the other attacks are blocked by some of the defense mechanisms or the patch of the vulnerability in question.

Our methodology leads to 4 categories of security metrics with respect to
vulnerabilities, defenses, threats, and situations.
As we review each category of security metrics,
we also discuss their theoretical and practical uses mentioned above
as well as what the {\em ideal metrics} may be, which hints at the gap between the state-of-the-art and the ideal metrics we need to close.
The insight behind the taxonomy is that, in principle, {\em situations} (or outcomes of cyber attack-defense interactions)
are caused by certain {\em threats} (or attacks) against systems that have certain {\em vulnerabilities} (including human factors) and employ certain {\em defenses}.
We here give a brief overview of the categories, which will be
respectively elaborated in Sections \ref{sec:knowing-your-vulnerabilities}-\ref{sec:knowing-the-situation}.

\subsubsection{Metrics for measuring vulnerabilities}
This category of metrics aim to measure the vulnerabilities of systems.
As illustrated in Figure \ref{fig:taxonomy}(a), an enterprise system consists of a vector $C(t)=(c_1(t),\ldots,c_n(t))$ of computers at time $t$.
As illustrated in Figure \ref{fig:taxonomy}(b), the enterprise system may have a vector $V(t)=(v_1(t),\ldots,v_n(t))$ of set of vulnerabilities at time $t$,
where $v_i(t)$ is, as illustrated in Figure \ref{fig:taxonomy-individual-view}, the vector of vulnerabilities with respect to $c_i(t)$.
Vulnerabilities include user vulnerabilities, password guessability, and software vulnerabilities.
Software vulnerabilities can be known or unknown (i.e., zero-day) to the defender.

\subsubsection{Metrics for measuring defenses}
This category of metrics aim to measure
the power or effectiveness of the defense mechanisms that are employed to protect enterprise and computer systems.
As Figure \ref{fig:taxonomy}(c) and Figure \ref{fig:taxonomy-individual-view} illustrate, we use blue bars to represent defenses and their thickness to indicate their power.
In practice,
some computers may be well defended (illustrated by thick blue bars), some computers may be poorly defended (illustrated by thin blue bars),
and some computers or zero-day vulnerabilities may not be defended at all (illustrated by the absence of blue bars).

\subsubsection{Metrics for measuring threats}
This category of metrics measure the threat landscape as well as the power or effectiveness of attacks.
The threat landscape describes aspects of the attacking computers.
As illustrated in Figure \ref{fig:taxonomy}(d) and Figure \ref{fig:taxonomy-individual-view},
we use red arrows to represent attacks and their {\em thickness} to indicate their attack power.
 Some computers may be attacked by powerful attacks (illustrated by thick arrows),
some computers may be attacked by less powerful attacks (illustrated by thin arrows), and some computers may not be attacked at all (illustrated by dash arrows).

\subsubsection{Metrics for measuring situations}
This category of metrics measure
outcomes of attack-defense interactions, especially the evolution of the global security state $S(t)$ over time $t$ \cite{SandersQUEST2011,XuCybersecurityDynamicsHotSoS2014}.
As illustrated in Figure \ref{fig:taxonomy}(e) and Figure \ref{fig:taxonomy-individual-view},
security state $s_i(t)=1$ means computer $c_i(t)$ is compromised at time $t$, and $s_i(t)=0$ otherwise.
However, the defender may not know the true state vector $S(t)=(s_1(t),\ldots,s_n(t))$ because of
 measurement or observation errors such as false-positives, false-negatives, and non-decisions, as illustrated in Figure \ref{fig:taxonomy}(f).
In other words, it is possible that the observation vector $O(t)=(o_1(t),\ldots,o_n(t))$ observed by the defender is not equal to $S(t)$.


\section{Metrics: Measuring System Vulnerabilities}
\label{sec:knowing-your-vulnerabilities}

These metrics aim to measure the vulnerabilities of enterprise and computer systems via
their users,
the passwords of their users,
their interfaces,
their software vulnerabilities, and
the vulnerabilities of the cryptographic keys they use.

\subsection{Measuring system users' vulnerabilities}

One metric is {\em user's susceptibility to phishing attacks} \cite{Sheng:2010:FPD:1753326.1753383}.
This online study of 1,001 users shows that phishing education can reduce the user's susceptibility to phishing attacks
and that young people (18 to 25 years old) are more susceptible to phishing attacks.
This metric is measured via the false-positive rate that a user treats legitimate email or website as a phish,
and the false-negative rate that a user treats a phishing email or website as legitimate and subsequently clicks the link in the email or submits information to the website.

Another metric is {\em user's susceptibility to malware infection} \cite{LalondeLevesque:2013:CSR:2508859.2516747}.
This clinical study of interactions between human users, anti-malware software, and malware
involves 50 users, who monitor their laptops for possible infections during a period of 4 months.
During this period of time, 38\% of users are found to be exposed to malware, which indicates the value of the
anti-malware tool (because these laptops would have been infected if anti-malware software was not used).
The study also shows that user demographics (e.g., gender, age) are {\em not} significant factors in determining a user's susceptibility to malware infection,
which contradicts the aforementioned finding in regards to users' susceptibility to phishing attacks \cite{Sheng:2010:FPD:1753326.1753383}.
Nevertheless, it is interesting to note that (i) users installing many applications are more susceptible to malware infections,
because the chance of installing malicious applications is higher, and
(ii) users visiting many websites are more susceptible to malware infections,
because some websites are malicious \cite{LalondeLevesque:2013:CSR:2508859.2516747}.

It is important to understand and measure
the degrees of users' susceptibilities to each individual class of attacks and to multiple classes of attacks collectively (e.g.,
multiple forms of social-engineering attacks).
For this purpose, research needs to be conducted to quantify
how the susceptibilities are dependent upon factors that affect users' security decisions (e.g., personality such as high vs. low attention control \cite{Neupane:2015:MNS:2810103.2813660}).
This area is little understood \cite{DBLP:conf/sp/HoweRRUB12,Sheng:2010:FPD:1753326.1753383,LalondeLevesque:2013:CSR:2508859.2516747},
but the reward is high.
For the theoretical use of security metrics,
these metrics can be incorporated into security models as parameters to model (e.g.) the time or effort that is needed in order for an attacker to
exploit user vulnerabilities to compromise a computer or to penetrate into an enterprise system.
For the practical use of security metrics,
these metrics can be used to tailor defenses for individual users (e.g.,
a careless employee may have to go through some security proxy in order to access Internet websites).
It would be appropriate to say that being able to measure these security metrics
is as important as being able to measure individual users' susceptibility to cancers because of (e.g.) her genes.
As the ability to quantify an individual's predisposition to diseases can  lead to proactive treatment, the ability to quantify security can lead to tailored and more effective defenses.

\subsection{Measuring password vulnerabilities}
The {\em parameterized password guessability} metric measures the number of guesses an attacker with a particular cracking algorithm
 (i.e., a particular threat model) needs to make before recovering a password
\cite{Weir:2010:TMP:1866307.1866327,Bonneau:2012:SGA:2310656.2310721,Kelley:2012:GAM:2310656.2310715,cracking:usenix15}.
This metric is easier to use
than earlier metrics such as {\em password entropy} \cite{NISTElectronicAuthenticationGuideline2006},
which cannot tell which passwords are easier
to crack than others, and {\em statistical password guessability} \cite{Bonneau:2012:SMI:2437647.2437657,Bonneau:2012:SGA:2310656.2310721,Kelley:2012:GAM:2310656.2310715},
which is more appropriate for evaluating passwords as a whole (rather than for evaluating them individually).

The parameterized password guessability metric should be used with caution if a single password cracking algorithm is used,
because different cracking algorithms can have very different strategies with varying results \cite{cracking:usenix15}.
When the defender is uncertain about the threat model, multiple cracking strategies need to be considered.
For both theoretical and practical uses of password vulnerability metrics,
we might need to consider the worst-case and/or the average-case parameterized password guessabilities.
This is one of the few sub-categories of security metrics that are better understood.

\subsection{Measuring interface-induced vulnerabilities}

The interface to access an enterprise or computer system from the outside world (e.g., service access points) offers potential opportunities for launching cyber attacks against the system.
The {\em attack surface} metric measures the number and severity of attack vectors that can be launched against a system through its service access points
such as sockets and RPC endpoints \cite{DBLP:journals/tse/ManadhataW11}.
It is worth mentioning that the attack surface is not necessarily dependent upon software vulnerabilities.
The attack surface should be used with caution because reducing the attack surface (e.g., uninstalling a security software) does not necessarily improve security \cite{TudorRAID2014}.
It has been suggested to define a variant of attack surface as the portion of the attack surface that has been exercised \cite{TudorRAID2014}.
This variant, while useful in analyzing historical data (i.e., incidents that have occurred), may or may not be appropriate
for measuring security in the future because an attack surface not exercised in the past may be exercised in the future.

Suppose we are to model security from higher levels of abstractions by treating system interface.
We would need to measure  {\em interface-induced system susceptibility}, which measures how the exercise of
attack surface is dependent upon the features of attack surfaces.
For practical purposes, it is ideal to be able to predict interface-induced system susceptibilities, namely the interfaces that will be exploited to launch attacks in the near future.
Knowing which interfaces are more likely to be abused to launch attacks would allow the defender to employ tailored defenses that pay particular attention
to these interfaces.


\subsection{Measuring software vulnerabilities}
Software vulnerabilities are the main venue for launching cyber attacks.
We classify the metrics for measuring software vulnerabilities into three sub-categories:
spatial characteristics, temporal characteristics, and severity.

\subsubsection{Measuring software vulnerability spatial characteristics}
These metrics reflect how spatially vulnerable an enterprise or computer system is.
The  number of {\em unpatched vulnerabilities} at time $t$ can be determined by using vulnerability scanners \cite{NIST800-55Rev1,CIS}.
The {\em vulnerability prevalence} metric measures the popularity of a vulnerability in a system \cite{Zhang:2014:AWK:2590296.2590300}.
This metric is important because a single vulnerability may exist in many computers of an enterprise or cloud system, and
because an attacker can launch a single attack against all the computers that possess a prevalent vulnerability.
Another variant metric is the number of {\em exploited vulnerabilities} that
have been exploited in the past \cite{TudorRAID2014,DBLP:conf/essos/Allodi15}.
This metric is important because
some vulnerabilities may never get exploited in the real world because, for example,
many vulnerabilities are difficult to exploit or their exploitation does not reward the attacker with many compromised computers.
For example, one study \cite{TudorRAID2014} shows that at most 35\% of the known vulnerabilities have been exploited, with
a small number of vulnerabilities (e.g., CVE-2008-4250 and CVE-2009-4324) being responsible for many attacks.
Another study \cite{DBLP:conf/essos/Allodi15} shows that 10\% of the vulnerabilities are responsible for 90\% attacks.

When using these metrics as parameters in security modeling, we would need to estimate the susceptibility
of a computer to attacks that exploit software vulnerabilities at time $t$.
When using these metrics to compare the security of two systems or the security of a system during two periods of time,
one must be cautious about (i) some vulnerabilities never being exploited, (ii) the varying capabilities of scanners in terms of their scanning depth and completeness, and (iii)
the threats may be different (e.g., two systems may be targeted by different attackers and there may be zero-day attacks that are not detected yet).
In other words, the theoretical and practical uses of these security metrics
require us to estimate, or even predict, the {\em vulnerability situation awareness} metric.
This metric measures
the number of vulnerabilities of a system at time $t$ and the likelihood of each of these vulnerabilities being exploited at time $t'\geq t$.

\subsubsection{Measuring software vulnerability temporal characteristics}

Temporal characteristics of software vulnerabilities include their evolution and lifetime.

\paragraph{Measuring evolution of software vulnerabilities}
The {\em historical vulnerability} metric measures the degree that a system is vulnerable, or the number of vulnerabilities, in the past \cite{Al-Shaer:2008:CON:1413140.1413189,4509855}.
The {\em future vulnerability} metric measures the number of vulnerabilities that will be discovered during a future period of time
\cite{Al-Shaer:2008:CON:1413140.1413189,4509855}.
Interesting variants of these metrics include historical exploited vulnerabilities, namely the number of vulnerabilities that were exploited in the past,
and future exploited vulnerabilities, namely the number of vulnerabilities that will be exploited during a future period of time.
The {\em tendency-to-be-exploited} metric measures the tendency that a vulnerability may be exploited,
where the ``tendency'' may be computed from (e.g.) the information that was posted on Twitter before vulnerability disclosures \cite{DBLP:conf/uss/SabottkeSD15}.
This metric may be used to prioritize vulnerabilities for patching.

\paragraph{Measuring software vulnerability lifetime}
It is ideal that each vulnerability is immediately patched upon its disclosure.
Despite the enforcement of patching policies, some vulnerabilities may never get patched.
The {\em vulnerability lifetime} metric measures how long it takes to patch a vulnerability since its disclosure.
Different vulnerability lifetimes may be exhibited at the client-end, the server-end, and the cloud-end.

Client-end vulnerabilities are often exploited to launch targeted attacks (e.g., spear-fishing) \cite{hardy2014targeted,DBLP:conf/uss/MarczakSMP14}.
These vulnerabilities are hard to patch completely
because of their prevalence (i.e., a vulnerability may appear in multiple programs)
 \cite{wineOakland2015}.
A study conducted in year 2010 \cite{FreiWP2010} shows that 50\% of 2 million Windows users in question are exposed to 297 vulnerabilities over a period of 12 months.
A more recent study \cite{wineOakland2015} shows that despite the presence of
13 automated patching mechanisms (other than the Windows update),
the median fraction of computers that are patched when exploits are available is no greater than 14\%,
the median time for patching 50\% of vulnerable computers is 45 days after disclosure.

One would think that server-end vulnerabilities are more rapidly patched than client-end ones.
Let us consider the disclosure of two severe vulnerabilities in OpenSSL.
First, for the pseudorandom-number-generation vulnerability in Debian Linux's OpenSSL,
a study \cite{Yilek:2009:PKP:1644893.1644896} shows that 30\% of the computers that were vulnerable  4 days after disclosure
remain vulnerable almost 180 days later (i.e., 184 days after disclosure).
This is somewhat surprising because
the private keys generated by the vulnerable computers might have been exposed to the attacker.
Second, for the Heartbleed vulnerability in
OpenSSL that can be remotely exploited to read
a vulnerable server's sensitive memory that may contain cryptographic keys and passwords,
a study \cite{Durumeric:2014:MH:2663716.2663755} estimates that 24\%-55\% of the HTTPS servers in Alexa's Top 1 Million websites were initially vulnerable.
Moreover, 11\% of the HTTPS servers in Alexa's Top 1 Million remain vulnerable 2 days after disclosure,
and 3\% of the HTTPS servers in Alexa's Top 1 Million were still vulnerable 60 days after disclosure.
%

One may think that vulnerabilities in the cloud are well managed, perhaps because
cloud users can run public virtual machine images (in addition to their own images).
A study \cite{Zhang:2014:AWK:2590296.2590300} shows that
many of the 6,000 public Amazon Machine Images (AMIs) offered by Amazon Web Services (AWS) Elastic Compute Cloud (EC2),
contain a considerable number of vulnerabilities, and that
Amazon typically notifies cloud users about vulnerabilities 14 days after their disclosure.

\medskip

Summarizing the temporal metrics discussed above,
we observe that defenders need to do a substantially better job at reducing the lifetime of software vulnerabilities after disclosure.
Because vulnerability lifetime may never be reduced to 0, it is important to know the vulnerability vector $V(t)$ or $v_i(t)$ at any time $t$.
For using vulnerability lifetime in security modeling, we need to know its statistical distribution and how the distribution is dependent upon various factors.



\subsubsection{Measuring software vulnerability severity}
This metric measures the degree of damage that can be caused by the exploitation of a vulnerability.
A popular example is the {\em CVSS score}, which considers the following three factors \cite{CVSS}.
The base score reflects the vulnerability's time- and environment-invariant characteristics, such as
its access condition, the complexity to exploiting it, and the impact once exploited.
The temporal and environmental scores reflect its time- and environment-dependent characteristics.
Another example is the {\em availability of exploits} in black markets \cite{Bilge:2012:BWK:2382196.2382284}, which
is interesting because the public release of vulnerabilities is often followed by the increase of exploits.

However, many vulnerabilities have the same CVSS scores \cite{Jansen09directionsin,DBLP:journals/tissec/AllodiM14}.
The practice of using CVSS scores (or base scores) to prioritize the patching of vulnerabilities
has been considered both harmful, because information about low-severity bugs can lead to the development of high-severity attacks
\cite{DBLP:journals/tissec/AllodiM14,Arnold:2009:SIR:1855568.1855584,Brumley:2008:APE:1397759.1398066},
and ineffective, because patching a vulnerability solely because of its high CVSS score makes no difference than patching vulnerabilities randomly
\cite{DBLP:journals/tissec/AllodiM14}.
For practical use, it would be ideal if we can precisely define the intuitive metric of {\em patching priority}.
For theoretical use, it would be ideal if we can quantify the {\em global damage} of a vulnerability to an enterprise system upon its exploitation,
which may in turn help measure the {\em patching priority}.

\subsection{Measuring cryptographic key vulnerabilities}

Cryptographic keys are vulnerable when the underlying random number generators are weak, as witnessed by
the pseudorandom-number-generation vulnerability in Debian Linux's OpenSSL \cite{Yilek:2009:PKP:1644893.1644896}.
Here we highlight the weak cryptographic keys caused by using
the fast {\tt /dev/urandom} as a replacement of the slow {\tt /dev/random} in Linux
\cite{Heninger:2012:MYP:2362793.2362828}.
The difference between them is that the former returns the requested number of bytes immediately even though not enough entropy
has been collected, while the latter returns the requested number of bytes only after the entropy pool contains the required fresh entropy.
As a consequence of using {\tt /dev/urandom}, the same key materials (e.g., prime numbers) can be generated and used by multiple devices.
A study
shows
that RSA private keys for 0.50\% of the TLS hosts examined and 0.03\% of
SSH hosts examined can be exposed because their RSA moduli shared nontrivial
factors \cite{Heninger:2012:MYP:2362793.2362828}.
The study also shows that the DSA private keys for 1.03\% of the SSH hosts examined can be extracted due to the insufficient randomness in their
digital signatures. These problems mainly exist in embedded devices, including routers and firewalls,
because they generate cryptographic keys on their first boot.

This kind of vulnerability should have been prevented by prudential engineering in the use of randomness,
which requires the programmer to understand, for example, the difference between {\tt /dev/random} and {\tt /dev/urandom}.
Nevertheless, it would be ideal to know whether a newly generated cryptographic key is weak or not.

\section{Metrics: Measuring Defenses}
\label{sec:knowing-your-defense}

These metrics measure the defenses employed to protect enterprise and computer systems via
the effectiveness of blacklisting,
the power of attack detection,
the effectiveness of software diversification,
the effectiveness of memory randomization, and
the overall effectiveness of these defenses.

\subsection{Measuring the effectiveness of blacklisting}
Blacklisting is a useful, lightweight defense mechanism.
Suppose a malicious entity (e.g., attacking computer, IP address,
malicious URL, botnet command-and-control server, and dropzone server) is observed at time $t$. Then, the traffic flowing to or from the malicious entity
can be blocked starting at some time $t'\geq t$.
The {\em reaction time} is the delay $t'-t$ between the observation of the malicious entity at time $t$
and the blacklisting of
the malicious entity at time $t'$ \cite{DBLP:conf/raid/KuhrerRH14}.
The {\em coverage} metric measures the portion of malicious entities that are blacklisted.
For example, a study
shows that the union of 15 malware blacklists covers only 20\% of the malicious domains that are compromised by some major malware families \cite{DBLP:conf/raid/KuhrerRH14}.

These metrics are with respect to the observers and blacklists in question.
For practical use, these metrics can be used to compare the effectiveness of different blacklists and can guide the design of better blacklisting solutions
(e.g., achieving a certain reaction time and a certain degree of coverage).
For theoretical use in security modeling, we might need to accommodate them into a unified metric, which may be called {\em blacklisting probability}
and may be measured by the conditional probability that a malicious entity at time $t$ (e.g., URL or IP address) is blacklisted at time $t$.
This would require us to understand the various factors that can impact malicious entities to be blacklisted.

\subsection{Measuring the power of attack detection}

Attack detection tools, such as cyber instruments (e.g., honeypots and blackholes that monitor unused IP addresses for attacks),
intrusion detection systems and anti-malware programs, aim to detect attacks. 
The effectiveness of attack detection can be measured by their individual effectiveness,
relative effectiveness, and collective effectiveness.

\subsubsection{Measuring the individual detection power}
For instrument-based attack detection, the {\em detection time} metric measures the delay between the time $t_0$ at which a compromised computer sends its first scan packet
and the time $t$ that a scan packet is observed by the instrument \cite{Rajab:2005:EDW:1251398.1251413}.
This metric depends on several factors, including malware spreading model, the distribution of vulnerable computers,
the size of the monitored IP address space, and the locations of the instrument.

For intrusion detection systems (including anomaly-based, host-based, and network-based),
the initial set of metrics for measuring their detection power are:
The {\em true-positive} rate, denoted by $\Pr(A|I)$, is defined as the probability that an intrusion ($I$) is detected as an alert that indicates attack ($A$).
The {\em false-negative} rate, denoted by $\Pr(\neg A|I)$, is defined as the probability that an intrusion is not detected as an attack.
The {\em true-negative} rate, denoted by $\Pr(\neg A|\neg I)$, is defined as the probability that a non-intrusion is not detected as an attack.
The {\em false-positive} rate, also called {\em false alarm rate}, denoted by $\Pr(A|\neg I)$, is defined as the probability that a non-intrusion is detected as an attack.
Note that $\Pr(A|I)+\Pr(\neg A|I)=\Pr(\neg A|\neg I)+\Pr(A|\neg I)=1$.
The {\em receiver operating characteristic} (ROC) curve reflects the dependence of the true-positive rate $\Pr(A|I)$
on the false-positive rate $\Pr(A|\neg I)$, and therefore can help determine the trade-off between the true-positive rate and the false-positive rate.

When using the preceding security metrics to compare the effectiveness of intrusion detection systems, care must be taken.
One issue is the event unit, such as packet vs. flow in the context of network-based intrusion detection \cite{GuAsiaCCS2006}.
Another issue is the base rate of intrusions, which can lead to misleading results if not adequately treated --- a phenomenon known as the {\em base-rate fallacy} \cite{AxelssonCCS09}.
In order to deal with the base-rate fallacy, one can
treat the input to an intrusion detection system as a stream ${\cal I}$ of 0/1 random variables (0 indicates benign or normal, 1 indicates malicious or abnormal),
and treat the output of the intrusion detection system as a stream ${\cal O}$ of 0/1 random variables (0 indicates no alert or normal, 1 indicates alert or abnormal).
Let $H({\cal I})$ and $H({\cal O})$ denote the entropy of ${\cal I}$ and ${\cal O}$, respectively.
The mutual information $I({\cal I},{\cal O})$ between ${\cal I}$ and ${\cal O}$, namely
$I({\cal I},{\cal O})=H({\cal I})-H({\cal I}|{\cal O})$, indicates the amount of uncertainty of ${\cal I}$ reduced after knowing ${\cal O}$.
The {\em intrusion detection capability} metric is defined as the normalization of $I({\cal I},{\cal O})$ with respect to $H({\cal I})$,
which reflects the base rate \cite{GuAsiaCCS2006}.

Intrusion detection may also be measured via the {\em cost} metric in the decision-theoretic framework \cite{GaffneyJrOakland2001}.
The cost includes both the operational cost of intrusion detection and the damage caused by false negatives.
Cardenas et al. \cite{BarasOakland2006} unified these metrics into a single framework of multi-criteria optimization,
which allows fair comparisons between intrusion detection systems in different operational environments.
We refer to a recent survey \cite{Milenkoski:2015:ECI:2808687.2808691} for more details.

\medskip

The metrics mentioned above are mainly geared towards the practical use of measuring the detection power of each individual detection system
and comparing the detection power of two detection systems.
When modeling intrusion detection systems as a component in a broader or holistic security model,
we may need to define and measure the {\em detection probability} metric as
the conditional probability that a compromised computer at time $t$ is also detected as compromised at time $t$,
namely $\Pr(o_i(t)=1|s_i(t)=1)$. This would require us to study how this probability is dependent upon other factors.

\subsubsection{Measuring the relative detection power}
This metric
\cite{boggs2011aldr,StolfoRAID2014Metrics}
reflects the effectiveness of a defense tool when employed in addition to other defense tools.
A defense tool does not offer any extra power if it cannot detect any attack that cannot be detected by the already deployed defense tools.
Let $\A$ denote a set of attacks, $\D=\{d_1,\ldots,d_n\}$ denote the set of $n$ defense tools, and
$X_d$ denote the set of attacks that are detected by defense tool $d\in \D$.
The relative power of defense tool $d'\in \D$ with respect to the set $D\subset \D$ of deployed defense tools
is defined as $\frac{|X_{d'}-\cup_{d\in D}X_d|}{|A|}$.

This kind of metric can be used to help decide whether to purchase/install a new defense tool or not, depending on its relative detection power.
This kind of metric could be used as parameters in security models that aim to characterize the global effectiveness of employing a defense tool.
It is worth mentioning that these metrics are measured based on attacks that have been seen in the past.
We might need to estimate these metrics  with respect to future attacks that may contain some unknown attacks,
which we may call {\em relative effectiveness against known and unknown attacks}.
This may require us to estimate the base rate of unknown attacks.

\subsubsection{Measuring the collective detection power}
This metric has been proposed for measuring the collective effectiveness of intrusion detection systems and
anti-malware programs \cite{boggs2011aldr,XuASEScienceJournal2012,StolfoRAID2014Metrics,mohaisen2014av,Yardon2014}.
Let $\A$ denote a set of attacks, $\D=\{d_1,\ldots,d_n\}$ denote the set of $n$ defense tools, $X_d$ denote the set of attacks that are detected by defense tool $d\in \D$.
The collective detection power of defense tools $D\subseteq \{d_1,\ldots,d_n\}$
is defined as
$\frac{|\cup_{d\in D}X_d|}{|A|}$ \cite{boggs2011aldr,StolfoRAID2014Metrics}.
For malware detection,
experiments \cite{XuASEScienceJournal2012,mohaisen2014av,Yardon2014} show that the collective use of multiple anti-malware programs still cannot detect all malware infections.
For example, one recent estimation \cite{Yardon2014} shows that anti-malware tools are only able to detect 45\% of attacks.

\medskip

The practical use of these metrics include the comparison of the collective effectiveness between two combinations of detection tools
and the evaluation of the effectiveness of defense-in-depth.
The theoretical use of these metrics include the incorporation of them as parameters into security models that aim to characterize the global
collective effectiveness of employing a combination of defense tools.
Like in the case of relative effectiveness mentioned above,
these metrics may need to be measured or estimated with respect to known and unknown attacks,
which we may call {\em collectiveness effectiveness against known and unknown attacks}.
This may also require us to estimate the base rate of unknown attacks.

\subsection{Measuring the effectiveness of Address Space Layout Randomization (ASLR)}
\label{sec:defense-power-ASLR}

Code injection  was a popular attack that aims to inject some malicious code into a running program and direct the processor to execute it.
The attack requires the presence of a memory region that is both executable and writable, which was possible because operating systems
 used to not distinguish programs and data.
The attack can be defeated by deploying Data Execution Prevention (DEP, also known as W$\oplus$X), which ensures
that a memory page can be writable or executable at any point in time, but not both.
The deployment of DEP made attackers move away from launching code injection attacks to launching code reuse attacks, which craft
attack payloads from pieces or ``gadgets'' of executable code that is already running in the system.
In order to launch a code-reuse attack, the attacker needs to know where to look for gadgets.
This was possible because the base addresses of code and data (including stack and heap) in the virtual memory used to be fixed.

One approach to defending against code reuse attacks is to use ASLR to ``blind'' the attacker, by randomizing the base addresses
(i.e., shuffling the code layout in the memory) such that the attacker cannot find useful gadgets.
Coarse-grained ASLR has the vulnerability that the leak or exposure of a single address gives the attacker adequate information to
extract all code addresses. Fine-grained ASLR do not suffer from this problem
(e.g., page-level randomization \cite{DBLP:conf/uss/BackesN14,Larsen:2014:SAS:2650286.2650803}),
but are still susceptible to attacks
that craft attack payloads from Just-In-Time (JIT) code \cite{Snow:2013:JCR:2497621.2498135}.
This attack can be defeated by {\em destructive code read}, namely that the code in executable memory pages is garbled once it is read \cite{Tang:2015:HTM:2810103.2813685}.
ASLR can also be enhanced by
preventing the leak of code pointers, while rendering leaks of other information (e.g., data pointers) useless for deriving code pointers \cite{Lu:2015:ASA:2810103.2813694}.



There are two metrics for measuring the effectiveness of ASLR.
One metric is the {\em entropy} of a memory section,
because a greater entropy would mean a greater effort in order for an attacker to compromise the system.
For example, a brute-force attack can feasibly defeat a low-entropy ASLR on 32-bit platforms \cite{Shacham:2004:EAR:1030083.1030124}.
The related {\em effective entropy} metric measures
the entropy in a memory section that the attacker cannot circumvent by exploiting the interactions between memory sections
\cite{herlands2014effective}.

\medskip

The two metrics mentioned above {\em indirectly} reflect the effectiveness of ASLR.
For practical use, we would need to measure the {\em direct} security gain offered by the deployment of ASLR
and/or the extra effort that is imposed on the attacker in order to circumvent ASLR.
Being able to measure the effectiveness of ASLR on individual computers, the resulting metrics
could be be incorporated into theoretical cyber security models to characterize their global effectiveness.

\subsection{Measuring the effectiveness of enforcing Control-Flow Integrity (CFI)}
\label{sec:defense-power-CFI}

Despite the employment of defenses such as the aforementioned DEP and ASLR, control-flow hijacking remains to be a big threat \cite{Szekeres:2013:SEW:2497621.2498101,Larsen:2014:SAS:2650286.2650803}.
Enforcing CFI has a great potential in assuring security.
The basic idea underlying CFI is to extract a program's Control-Flow Graph (CFG), typically from its source code via static analysis, and
then instrument the corresponding binary code to abide by the CFG at runtime.
This can be implemented by runtime checking of the tags that were assigned to the indirect branches in the CFG, such as indirect calls, indirect
jumps, and returns.
Since it is expensive to enforce CFI according to the entire CFG \cite{Abadi:2005:CI:1102120.1102165},
practical solutions enforce weaker restrictions via a limited number of tags.
It was known that coarse-grained enforcement of CFI uses a small number of tags and can be compromised by
code reuse attacks \cite{Goktas:2014:OCO:2650286.2650770,Davi:2014:SGI:2671225.2671251,Carlini:2014:RSD:2671225.2671250}.
This inadequacy led to fine-grained CFI, such as
the enforcement of forward-edge control integrity (i.e. indirect calls but not returns) \cite{Tice:2014:EFC:2671225.2671285}, and
the use of message authentication code to prevents unintended control transfers in the CFG \cite{Mashtizadeh:2015:CCE:2810103.2813676}.
Even a fully accurate, fine-grained CFG can be compromised by the {\em control flow bending} attack \cite{DBLP:conf/uss/CarliniBPWG15},
which however can be mitigated by per-input CFI \cite{Niu:2015:PCI:2810103.2813644}.

How should we measure the power of CFI?
First, the power of CFI is fundamentally limited by the {\em accuracy} of the CFGs \cite{Evans:2015:CJW:2810103.2813646}.
Because CFGs are computed via static analysis, their accuracy depends on
{\em sound} and {\em complete} pointer analysis, which is undecidable in general \cite{Ramalingam:1994:UA:186025.186041}.
The trade-off of using an unsound pointer analysis is that all of the due connections may not be reported and therefore can cause false-positives.
The trade-off of using an incomplete pointer analysis is that excessive connections may be reported (i.e., over-approximation),
which can be exploited to run arbitrary code despite the enforced fine-grained CFI \cite{Evans:2015:CJW:2810103.2813646}.
Second, we need to measure the {\em resilience} of a CFI scheme against control flow bending attacks,
which ideally reflects the effort (or premises) that an attacker must make (or satisfy) in order to evade the CFI scheme.
This metric would allow us to compare the resilience of two CFI schemes.
Third, it would be ideal if we can measure the power of CFI via the classes of attacks that it can defeat.
The key issue here is to have a formalism by which we can {\em precisely} classify attacks.
The challenge is that there could be infinitely many attacks and it is not clear what would be the right formalism.

\subsection{Measuring overall defense power}
In the above we discussed the defense power of individual defense mechanisms.
In practice, different kinds of defense mechanisms are used together and therefore we need to consider the overall defense power.
This is a field that is much less understood.
The well known approach is to use penetration test to evaluate an enterprise or computer system's
{\em penetration resistance}. This metric can be measured as the
effort (e.g., person-day or cost) it requires for the red team to penetrate into the system \cite{DBLP:conf/discex/Levin03}.
This metric can be used to compare the effectiveness of two systems against the same red team.
On the other hand, there have been some initial studies at measuring the power of Moving Target Defense (MTD), which may
deploy a set of MTD mechanisms together.
The analytic approach aims to indirectly measure the degree that an enterprise system can tolerate
some undesirable security configurations, under which the global security state $S(t)$ may evolve towards many computers are compromised \cite{DBLP:conf/hotsos/HanLX14}.
Complementary to the analysis approach, there have been proposals for evaluating the effectiveness of MTD via
experimentation \cite{Zaffarano:2015:QFM:2808475.2808476}, emulation \cite{Eskridge:2015:VCE:2808475.2808486}, and simulation \cite{Prakash:2015:EGA:2808475.2808483}.

When using these metrics, one must be cautious about what the penetration resistance is with respect to the specific red team,
which may reassemble real-world attackers to a certain extent.
Moreover, one should bear in mind that the identification of security holes  does not offer any quantitative security \cite{SandersSP14}.
For theoretical and practical uses, it is ideal that
the {\em penetration resistance} should at least consider some
hypothetical new attacks, which may exploit some known or zero-day vulnerabilities (i.e., ``what if'' analysis).
Moreover, systematic metrics need to be devised to directly measure the effectiveness of MTD,
which may be in terms of some {\em direct effectiveness} metrics,
such as the security gained by launching MTD and/or the extra effort imposed on the attacker in order to achieve attack goals.

\section{Metrics: Measuring Threats}
\label{sec:knowing-your-enemy}

These metrics measure the threats against an enterprise or computer system via
the threat landscape,
the threat of zero-day attacks,
the power of individual attacks,
the sophistication of obfuscation, and
the evasion capability.

\subsection{Measuring the threat landscape}

The threat landscape can be characterized via multiple attributes.
One attribute is the attack vector.
The number of {\em exploit kits} metric describes the number of automated attack tools
that are available in the black market \cite{RandReportOnBlackMarket2014}.
This metric can be extended to accommodate, for example, the vulnerabilities that the exploits are geared for.
This is a good indicator of cyber threats because most attacks
would be launched from these exploit kits \cite{TudorRAID2014,DBLP:conf/essos/Allodi15}.

The {\em network maliciousness} metric \cite{BaileyNDSS2014} measures the fraction of blacklisted IP addresses in a network.
The study \cite{BaileyNDSS2014}  shows that there were 350 autonomous systems which had at least 50\% of their IP addresses blacklisted.
Moreover, there was a correlation between mismanaged networks and malicious networks, where
``mismanaged networks'' are those networks that do not follow accepted policies/guidelines.
The related {\em rogue network} metric measures the population of networks that were
abused to launch drive-by download or phishing attacks \cite{KirdaACSAC09}.
The {\em ISP badness} metric quantifies the effect of spam from one ISP or Autonomous System (AS) on the rest of the Internet \cite{DBLP:conf/fc/JohnsonCGC12}.
The {\em control-plane reputation} metric quantifies the maliciousness of attacker-owned (i.e., rather than legitimate but mismanaged/abused)
ASs based on their control plane information (e.g., routing behavior),
which can achieve an {\em early-detection time} of 50-60 days (before these malicious ASs are noticed by other defense means) \cite{Konte:2015:ARS:2785956.2787494}.
Malicious, rogue, and bad networks, once detected, can be filtered by enterprise systems via blacklisting.

The {\em cybersecurity posture} metric measures the dynamic threat imposed by the attacking computers \cite{XuInTrust2014}.
It may include the attacks observed at honeypots, network telescopes, and/or production enterprise systems.
One related metric, the {\em sweep-time}, measures the time it takes for each computer or IP address in a target enterprise system to be scanned or attacked at least once  \cite{XuInTrust2014}.
Another related {\em attack rate} metric measures the number of attacks that arrive at a system of interest per unit time \cite{XuTIFS2013,XuTIFS2015}.
These metrics reflect the aggressiveness of cyber attacks.

Although the security metrics mentioned above can reflect some aspects of the threat landscape,
we might need to define what may be called {\em comprehensive cyber threat posture}, which reflects the holistic threat landscape.
This metric is useful because the threat landscape could be used as the ``base rate'' (in the language of intrusion detection systems) that
can help fairly compare the overall defense effectiveness of two enterprise systems.
It is interesting to investigate how these security metrics should be incorporated into security models as parameters for analyzing,
for example, the evolution of the global security state $S(t)$ over time $t$.

\subsection{Measuring zero-day attacks}

The number of {\em zero-day attacks} measures how many zero-day attacks were launched during a past period of time.
For example, Symantec estimated that there were 8-15 zero-day vulnerabilities between 2006 and 2011,
among which 9 were exploited to launch zero-day attacks in 2008,
12 were exploited to launch zero-day attacks in 2009,
14 were exploited to launch zero-day attacks in 2010, and
8  were exploited to launch zero-day attacks in 2011 \cite{Symantec-0-day-report-April-2012}.
The {\em lifetime of zero-day attacks} measures the period of time between when the attack was launched and when
the corresponding vulnerability is disclosed to the public.
One study shows that the lifetime of zero-day attacks
can last 19-900 days, with a median of 240 days and an average of 312 days \cite{Bilge:2012:BWK:2382196.2382284}.
The  number of {\em zero-day attack victims} measures the number of computers that were compromised by zero-day attacks.
It has been reported that most zero-day attacks affected a small number of computers, with a few exceptions (e.g., Stuxnet) \cite{Bilge:2012:BWK:2382196.2382284}.
This suggests that zero-day attacks are mainly used for launching targeted attacks.

The preceding metrics are typically measured after the fact.
For both theoretical and practical uses of security metrics measuring zero-day attacks, one important metric that has yet to be studied
is the {\em susceptibility of a computer to zero-day attacks}. Metrics of this kind, once understood, not only could help
allocate defense resources to carefully monitor the computers that are more susceptible to zero-day attacks,
but also could be incorporated into security models to analyze the evolution of the global security state $S(t)$ over time $t$.

\subsection{Measuring the attack power}

Now, we discuss the metrics for measuring the power of targeted attacks,
the power of botnets,
the power of malware spreading, and
the power of attacks that exploit multiple vulnerabilities in enterprise systems.

\subsubsection{Measuring the power of targeted attacks}

The success of targeted attacks or Advanced Persistent Threats (APT) often depends on the delivery of malware and the tactics
that are used to lure the target to open malicious email attachments.
Let $\alpha$ denote the social engineering tactics, ranging from the least sophisticated to the most sophisticated (e.g., $\alpha\in \{0,\ldots,10\}$).
Let $\beta$ denote the technical sophistication of the malware that are used in the attacks,
also ranging from the least sophisticated to the most sophisticated (e.g., $\beta\in [0,1]$).
The {\em targeted threat index} metric, which measures the level of
targeted malware attacks, can be defined as $\alpha \times \beta$ \cite{hardy2014targeted}.

The preceding metric represents a first step in measuring the power of targeted attacks, and much research has yet to be done.
It would be ideal if we can measure the {\em susceptibility of a computer to targeted attacks}, for which
the user's susceptibility to social-engineering attacks metric mentioned above would be an integral factor.
Being able to measure metrics of this kind allows the defender to tailor defenses for the computers that are more vulnerable to targeted attacks.
These metrics could also be immediately incorporated into security models that analyze (e.g.) the evolution of $S(t)$ over time $t$.

\subsubsection{Measuring the attack power of botnets}
The threat of botnets can be characterized by several metrics.
The first metric is {\em botnet size}.
It is natural to count the number $x$ of bots belonging to a botnet.
It is important to count the number of bots that can be instructed to launch attacks
(e.g., distributed denial-of-service attacks) at a point in time $t$, denoted by $y(t)$.
Due to factors such as the diurnal effect, which explains why some bot computers are powered off during night hours at local time zones,
$y(t)$ is often much smaller than $x$ \cite{DagonNDSS06}.
A related metric is the network bandwidth that a botnet can use to launch denial-of-service attacks \cite{DBLP:conf/acsac/DagonGLL07}.
The second metric is {\em botnet efficiency}, which can be defined as the network diameter of
the botnet network topology \cite{DBLP:conf/acsac/DagonGLL07}.
This metric measures a botnet's capability in communicating command-and-control messages and updating bot programs.
The third metric is {\em botnet robustness}, which measures the robustness of botnets under random or intelligent disruptions \cite{DBLP:conf/acsac/DagonGLL07}.
There has been a body of literature \cite{AlbertReview02} on measuring complex network robustness that can be adopted for characterizing botnets.

Although the above metrics measure botnets from some intuitive aspects,
it remains elusive to define the intuitive metric of {\em botnet attack power}, which is important because it can prioritize
the countermeasures against botnets.
Moreover, the intuitive {\em botnet resilience} metric would need to take into consideration the counter-countermeasures that may be
employed by the attacker during the process that the defender launches countermeasures against botnets.

\subsubsection{Measuring malware spreading power}

The {\em infection rate} metric, denoted by $\gamma$, measures the average number of vulnerable computers that are infected by
a compromised computer (per time unit) at the early stage of spreading \cite{DBLP:conf/infocom/ChenJ07}.
Intuitively, $\gamma$ depends on the scanning strategy. Here we only mention the random scanning strategy.
With random scanning, the malware scans over the vulnerable computers uniformly at random.
Denote by $z$ the number of scans and infections that are made by an infectious computer (per time unit).
Denote by $w$ the number of vulnerable computers. For the IPv4 address space, the infection rate is $\gamma=zw/2^{32}$ \cite{DBLP:conf/infocom/ChenJ07}.

The infection rate metric can be used to compare the infection rate of different malware and has been used in
various models (e.g., \cite{DBLP:conf/infocom/ChenJ07}) to derive interesting results.
It should be noted that the infection rate is defined with respect to the early stage of infection.
We would need other metrics to measure, such as the {\em sweep-time} that all or a fraction of the vulnerable computers that
will be infected. This metric is hard to compute for arbitrary scanning strategies.
Another interesting metric is the number of {\em wasted scans}, namely those which arrive at some infected computers.
This metric reflects the unstealthly nature of malware spreading.

\subsubsection{Measuring the power of attacks that exploit multiple vulnerabilities}
Vulnerabilities can be exploited in a chaining fashion.
There is a large body of literature in this field, including
{\em attack graphs}  (see, e.g.,
\cite{Phillips:1998:GSN:310889.310919,Ritchey:2000:UMC:882494.884423,Sheyner:2002:AGA:829514.830526,Jha:2002:TFA:794201.795177,Ammann:2002:SGN:586110.586140,JajodiaDSN2012,%
Homer:2013:AVM:2590624.2590627,OuCyberSA2014}),
{\em attack trees} \cite{Schneier:2000:SLD:517959}, and {\em privilege trees} \cite{DacierIFIPSEC2006,OrtaloIEEETSE2009}.
At a high level, these models
accommodate system vulnerabilities,
vulnerability dependencies (i.e., prerequisites), firewall rules, etc.
In these models, the attacker is initially in some security state
and attempts to move from the initial state to some goal state, which often corresponds to the compromise of computers.
 These studies have led to a rich set of metrics, such as the following.

The {\em necessary defense} metric measures the minimal set of defense countermeasures that must be employed in order to thwart a certain attack \cite{Sheyner:2002:AGA:829514.830526}.
The greater the necessary defense, the more powerful the attack.

The {\em weakest adversary} metric measures the minimum adversary capabilities that are needed in order to achieve an attack goal \cite{PamulaQoP2006}.
This metric can be used to compare the power of two attacks with respect to some attack goal(s).  For example,
one attack has the required weakest adversary capabilities, but the other does not.

The {\em existence, number, and lengths of attack paths} metrics measure
the these attributes of attack paths from an initial state to the goal
state \cite{Ritchey:2000:UMC:882494.884423,Sheyner:2002:AGA:829514.830526,Jha:2002:TFA:794201.795177,OuCyberSA2014}.
These metrics can be used to compare two attacks. For example, the attack that has a set $X$ of attack paths is more powerful than
another attack that has a set $Y$ of attack paths, where $Y \subset X$.

The {\em $k$-zero-day-safety} metric measures the number of zero-day vulnerabilities that are needed in order for an attacker to compromise a target \cite{WangESORICS08}.
This metric can be used to compare the power of two attacks as follows:
An attack that requires $k_1$ zero-day vulnerabilities in order to compromise a target is more powerful that an attack
that requires $k_2$ zero-day vulnerabilities, where $k_1< k_2$.

The {\em effort-to-security-failure} metric measures what the attacker needs to do in order to move from an initial set of privileges to the goal set of escalated privileges
\cite{DacierIFIPSEC2006,OrtaloIEEETSE2009}.
An attack that incurs a smaller effort-to-security-failure is more powerful than an attack that requires a greater effort, assuming the efforts are comparable.

Although the metrics mentioned are useful, it would be ideal if we can measure what we call {\em multi-stage attack power},
which may be able to incorporate all the metrics mentioned above into a single one.
This metric could also be incorporated into security models to analyze (e.g.) the evolution of security state $S(t)$ over $t$.
One barrier is to systematically treat unknown vulnerabilities.

\subsubsection{Measuring the power of evasion against learning-based detection}
Sophisticated attacks can evade the defense system by, for example, manipulating some features that are used in the detection models (e.g., classifiers).
This problem is generally known as {\em adversarial machine learning} \cite{Washington:Nilesh,Lowd:2005:AL:1081870.1081950,Huang:2011:AML:2046684.2046692,Srndic:2014:PEL:2650286.2650798}.
There is a spectrum of evasion scenarios, which vary in terms of the information the attacker knows about the detection models,
such as (i) knowing only the feature set used by the defender,
(ii) knowing both the feature set and the training samples used by the defender,
and (iii) knowing the feature set, the training samples, and the attack detection model (e.g., classifiers) used by the defender \cite{Srndic:2014:PEL:2650286.2650798,XuIEEECNS2014}.
The evaluation of effectiveness is typically based on metrics such as false-positive and false-negative rates as a consequence of
applying a certain evasion method.

It is ideal if we can measure the {\em evasion capability} of attacks.
This not only allows us to compare the evasion power of two attacks,
but also can possibly be used to compute the damage that can be caused by evasion attacks.
Despite the many efforts (cf. \cite{Srndic:2014:PEL:2650286.2650798} for extensive references),
this aspect of security is far from understood.

\subsection{Measuring obfuscation sophistication}
Obfuscation based on tools such as run-time packers have been widely used by malware-writers to defeat static analysis.
Despite the numerous studies that have been surveyed elsewhere \cite{Roundy:2013:BOP:2522968.2522972,DBLP:conf/sp/Ugarte-PedreroB15},
we understand very little about how to quantify the obfuscation capability of malware.
Nevertheless, there have been some notable initial efforts.
The {\em obfuscation prevalence} metric measures the occurrence of obfuscation in malware samples \cite{Roundy:2013:BOP:2522968.2522972}.
The {\em structural complexity} metric measures the runtime complexity of packers in terms of their
layers, granularity etc. \cite{DBLP:conf/sp/Ugarte-PedreroB15}.

It is ideal if we can measure the {\em obfuscation sophistication} of a malware, perhaps in terms of
the amount of effort that is necessary for unpacking the packed malware.
One practical use is to automatically differentiate the malware samples that must be manually unpacked from those that can be automatically unpacked.
One possible theoretical use is to incorporate it as a parameter in a model for analyzing the evolution of security state $S(t)$ over time $t$.

\section{Metrics: Measuring Situations}
\label{sec:knowing-the-situation}

Situation is the comprehensive manifestation of attack-defense interactions with respect to an enterprise or computer system.
These metrics measure situations via security states, security incidents, and security investments.

\subsection{Measuring the evolution of security state}

As illustrated in Figures \ref{fig:taxonomy}-\ref{fig:taxonomy-individual-view}, the security state vector of an enterprise system $S(t)=(s_1(t),\ldots,s_n(t))$ and
the security state $s_i(t)$ of computer $c_i(t)$ both dynamically evolve as a outcome of attack-defense interactions.
These metrics aim to measure the dynamic security states.
However, the measurement process often incurs errors, such as false-positives and false-negatives.
As a consequence, the observed state $O(t)=(o_1(t),\ldots,o_n(t))$ is often different from the true state $S(t)$.
The {\em fraction of compromised computers} is $|\{i: i\in \{1,\ldots, n\} \wedge s_i(t)=1\}|/n$.
It has been shown that under certain circumstances,
there can be some fundamental connection between the global security state and a very small number of nodes that can be monitored carefully \cite{XuTAAS2012}.
An alternative metric is the {\em probability a computer is compromised at time $t$}, namely $\Pr[s_i(t)=1]$ as illustrated in Figure \ref{fig:taxonomy}.
This metric has been proposed in some recent studies that aim to quantify the security in enterprise systems
(e.g., \cite{SandersQUEST2011,DBLP:journals/im/XuX12,DBLP:conf/hotsos/DaXX14,XuCybersecurityDynamicsHotSoS2014,XuHotSoS15,XuInternetMath2015ACD,XuInternetMath2015Dependence,XuEmergentBehaviorHotSoS2014,XuHotSOS14-MTD,XuTAAS2014,XuGameSec13,XuIEEETDSC2012-multivirus,XuTAAS2012,XuTDSC2011}).
These studies represent some early-stage investigations towards modeling security from a holistic perspective.

Knowing the dynamic security state can help the defender make the right decision.
For example, knowing the probabilities that computers are compromised at time $t$, namely $\Pr[s_i(t)=1]$ for every $i$,
allows the defender to use an appropriate
threshold cryptographic mechanism \cite{DF89} to tolerate the compromises.
However, faithful security models may require to accommodate many, if not all, of the aforementioned metrics as parameters.
Moreover, it is important to know $S(t)$ and $s_i(t)$ for any $t$, rather than for $t\to\infty$.
These impose many open problems that remain to be investigated \cite{XuCybersecurityDynamicsHotSoS2014}.

\subsection{Measuring security incidents}

\subsubsection{Measuring spatial characteristics of security incidents}

Spatial characteristics of incidents can be described by the {\em incident rate} metric \cite{CIS},
which measures the fraction of computers that are successfully infected or attacked at least once during a period of time.
The incident rate is often smaller than the {\em encounter rate}, which measures the fraction of computers that encountered some malware or attack during a period of time
\cite{DBLP:conf/ccs/YenHORJ14,Mezzour:2015:ESG:2746194.2746202}.
This is because some malware encounters and attacks are defeated by the deployed defense, which can be measured through the {\em blocking rate} metric,
namely the difference between the {\em encounter rate} and the {\em incident rate}.
For example, Microsoft reports 21.2\% and 19.2\% of the reporting computers in question encountered malware respectively in 2013 and 2014,
and malware infection rates are much less than malware encounter rates
\cite{MicrosoftIntelligenceReport}.
Another study based on the anti-virus software reports from an enterprise of 62,884 computers shows a 15.31\% malware encounter rate
during a period of 4 months \cite{DBLP:conf/ccs/YenHORJ14}.

These metrics may be used as alternative to the global security state $S(t)$, especially when $S(t)$ is difficult to obtain for arbitrary $t$
(rather than for $t\to \infty$).
It would be ideal if we can measure the {\em incident occurrence frequency} as an approximation of the number of compromised computers,
namely $|\{i: i\in \{1,\ldots, n\} \wedge s_i(t)=1\}|$, which may be represented as some mathematical functions
of system features. A recent study shows the possibility of predicting data breaches from symptoms of the network in question
(e.g., network mismanagement and blacklisted IP addresses) \cite{DBLP:conf/uss/LiuSZNKBL15}.
One should be cautions when using these metrics to compare the security of two systems,
because they may be attacked with different attack vectors.

\subsubsection{Measuring temporal characteristics of security incidents}

The temporal characteristic of incidents can be described by the {\em delay in incident detection} \cite{CIS},
which measures the time between when an incident occurred and when the incident is discovered.
Another metric is the {\em time between incidents} \cite{CIS,HolmIEEETDSC2014}, which measures the period of time between two incidents.
Yet another metric is the {\em time-to-first-compromise} metric \cite{JonssonIEEETSE1997,TrivediDSN2002,HolmIEEETDSC2014},
which measures the duration of time between a computer starts to run and the first malware alarm is triggered on the computer,
(alarm indicating detection rather than infection).
A study based on a dataset of 5,602,097 malware alarms, which correspond to 203,025 malware attacks against 261,757 computers between 10/15/2009 and 8/10/2012,
shows that the time-to-first-compromise follows the Pareto distribution \cite{HolmIEEETDSC2014}.

These metrics may be used as alternative to the global security state $S(t)$, especially when $S(t)$ is difficult to predict for arbitrary $t$ (rather than for $t\to \infty$).
It would be ideal if we can predict the {\em incident occurrence frequency} as an approximation of the number of compromised computers at a future time $t$,
namely $|\{i: i\in \{1,\ldots, n\} \wedge s_i(t)=1\}|$.
One should be cautions when using these metrics to compare the security of a system during two different periods of time,
because the threats would be different.

\subsubsection{Measuring the damage of security incidents}

The damage caused by incidents can be measured by
the {\em cost of incidents} metric \cite{CIS}, which measures
the count or cost of detected security incidents (i.e., successful infections or attacks) that have occurred by in a system during a period of time.
The cost of incidents may include both the direct cost (e.g., the amount of lost money) and the indirect cost (e.g., negative publicity and/or the recovery cost).
A very recent survey shows that the remediation cost per insider attack is \$500,000 \cite{Infosecbuddy2015}.

These metrics are useful, but we understand these metrics very little.
It would be ideal if we can predict the cost of incidents that occur in a future period of time, which
may depend on factors such as the {\em delay in detection} metric mentioned above.

\subsection{Measuring security investment}

These metrics include {\em security spending} as percentage of IT budget \cite{NIST800-55Rev1,CIS} and {\em security budget} allocations \cite{CIS}.
The former is important because enterprises want to know whether their security expenditure is justified by the
security performance and is comparable to other organizations' security investment.
The latter indicates how the security budget is allocated for various security activities and resources.

It is important to understand the {\em payoff of security investment}, which requires to investigate
whether or not there is some inherent relationship between the cost of security incidents and
factors such as the delay in the detection of security incidents, which may depend on security investment.

\section{Discussion}
\label{sec:discussion}

\input{open-problems.tex}

\section{Conclusion}
\label{sec:conclusion}

We have presented a survey of security metrics.
The survey is centered on the insight that ``situations (or outcomes of cyber attack-defense interactions)
are caused by certain threats (or attacks) against systems that have certain vulnerabilities (including human factors) and employ certain defenses.''
The resulting taxonomy contains 4 categories of security metrics:
those for measuring the system {\em vulnerabilities},
those for measuring the {\em defenses},
those for measuring the {\em threats},
and those for measuring the {\em situations}.
In addition to systematically reviewing the security metrics that have been proposed in the literature, we propose
what we believe to be desirable security metrics.
We discuss the gaps between the state-of-the-art metrics and the desirable metrics.
We also discussed some fundamental problems that must be adequately addressed in order to bridge these gaps,
including what academia, industry and government should do in order to catalyze the research in security metrics.

\bibliographystyle{ACM-Reference-Format-Journals}

\bibliography{metrics-bib,mtd,crypto,complex-network,measurement,botnet}

\end{document}

%% file: abstract.tex
The importance of security metrics can hardly be overstated.
Despite the attention that has been paid by the academia, government and industry in the past decades,
this important problem stubbornly remains open.
In this survey, we present a survey of knowledge on security metrics. The survey is centered on a novel
taxonomy, which classifies security metrics into four categories:
metrics for measuring the system {\em vulnerabilities},
metrics for measuring the {\em defenses},
metrics for measuring the {\em threats},
and metrics for measuring the {\em situations}.
The insight underlying the taxonomy is that situations (or outcomes of cyber attack-defense interactions)
are caused by certain threats (or attacks) against systems that have certain vulnerabilities (including human factors) and employ certain defenses.
In addition to systematically reviewing the security metrics that have been proposed in the literature, we  discuss
the gaps between the state of the art and the ultimate goals.
 

%% file: open-problems.tex
Table \ref{table:summary-of-metrics} summarizes the taxonomy and security metrics systemized above.
The column of {\em desirable security metrics} shows that there are big gaps between the state-of-the-art and the ultimate goals.
Now we discuss some fundamental questions that must be addressed in order to bridge these gaps.

\begin{sidewaystable}
\centering
\resizebox{6in}{!}{
\begin{minipage}{\textwidth}
\tbl{Summary of the taxonomy, the representative examples of security metrics systemized in the present paper, and the desirable metrics that are discussed in the text ($^\dag$ indicates that the desirable metric is little understood).``(Avoidable via prudential engineering)'' means the attacks in question can be avoided by prudential engineering.}{
\begin{tabular}{|l|p{0.7\textwidth}|p{0.3\textwidth}|}
\hline
Measuring what? &  representatives of metrics systemized in the paper & desirable security metrics \\
\hline\hline
\multicolumn{3}{|l|}{Measuring system vulnerabilities} \\ \hline
~~users' vulnerabilities & user's susceptibility to phishing attacks  \cite{Sheng:2010:FPD:1753326.1753383}, user's susceptibility to malware infection \cite{LalondeLevesque:2013:CSR:2508859.2516747} & user's susceptibility to class(es) of attacks (e.g., social-engineering)$^\dag$\\
~~password vulnerabilities & parameterized/statistical password guessability \cite{Weir:2010:TMP:1866307.1866327,Bonneau:2012:SGA:2310656.2310721,Kelley:2012:GAM:2310656.2310715,cracking:usenix15,Bonneau:2012:SMI:2437647.2437657}, password entropy \cite{NISTElectronicAuthenticationGuideline2006} & worst-case and average-case parameterized password guessability \\
~~system interface & attack surface \cite{DBLP:journals/tse/ManadhataW11}, exercised attack-surface \cite{TudorRAID2014} & interface-induced susceptibility$^\dag$ \\
~~software vulnerabilities & &\\
~~~~vulnerability spatial characteristics & unpatched vulnerabilities \cite{NIST800-55Rev1,CIS}, exploited vulnerabilities \cite{TudorRAID2014,DBLP:conf/essos/Allodi15}, vulnerability prevalence  \cite{Zhang:2014:AWK:2590296.2590300} & vulnerability situation awareness$^\dag$ \\
~~~~vulnerability temporal characteristics & historical (exploited) vulnerability \cite{Al-Shaer:2008:CON:1413140.1413189,4509855}, future (exploited) vulnerability \cite{Al-Shaer:2008:CON:1413140.1413189,4509855}, tendency-to-be-exploited \cite{DBLP:conf/uss/SabottkeSD15}, vulnerability lifetime \cite{FreiWP2010,wineOakland2015,Yilek:2009:PKP:1644893.1644896,Durumeric:2014:MH:2663716.2663755,Zhang:2014:AWK:2590296.2590300},  & vulnerability vector at any time$^\dag$, distribution of vulnerability lifetime$^\dag$ \\
~~~~vulnerability severity & CVSS score \cite{CVSS}, availability of exploit \cite{Bilge:2012:BWK:2382196.2382284} & patching priority$^\dag$, global damage$^\dag$ \\
~~cryptographic key vulnerabilities &  vulnerable cryptographic keys \cite{Yilek:2009:PKP:1644893.1644896,Heninger:2012:MYP:2362793.2362828,Durumeric:2014:MH:2663716.2663755} & (avoidable via prudential engineering) \\ \hline\hline
\multicolumn{3}{|l|}{Measuring defenses}\\ \hline
~~effectiveness of blacklisting & reaction time \cite{DBLP:conf/raid/KuhrerRH14}, coverage \cite{DBLP:conf/raid/KuhrerRH14} & blacklisting probability$^\dag$ \\
~~attack detection power & &\\
~~~~individual detection power & detection time \cite{Rajab:2005:EDW:1251398.1251413}, false-positive, false-negative, true-positive, true-negative, ROC, intrusion detection capability \cite{GuAsiaCCS2006}, cost \cite{GaffneyJrOakland2001} & detection probability$^{\dag}$ \\
~~~~relative detection power  & relative effectiveness \cite{boggs2011aldr,StolfoRAID2014Metrics} & relative effectiveness against unknown attacks$^\dag$ \\
~~~~collective detection power & collective effectiveness  \cite{boggs2011aldr,XuASEScienceJournal2012,StolfoRAID2014Metrics,mohaisen2014av,Yardon2014} & collective effectiveness against unknown attacks$^\dag$ \\
~~ASLR effectiveness & entropy  \cite{Shacham:2004:EAR:1030083.1030124}, effective entropy \cite{herlands2014effective} & security gain$^\dag$, extra attack effort$^\dag$ \\
~~CFI effectiveness & CFG accuracy \cite{Evans:2015:CJW:2810103.2813646}, & CFI resilience$^\dag$, CFI power$^\dag$ \\
~~overall defense power & penetration resistance \cite{DBLP:conf/discex/Levin03}, indirect MTD effectiveness \cite{DBLP:conf/hotsos/HanLX14} & resistance against unknown attacks$^\dag$, direct MTD effectiveness$^\dag$ \\\hline\hline
\multicolumn{3}{|l|}{Measuring threats} \\ \hline
~~threat landscape & exploit kits \cite{RandReportOnBlackMarket2014}, malicious network \cite{BaileyNDSS2014}, rogue network \cite{KirdaACSAC09}, ISP badness \cite{DBLP:conf/fc/JohnsonCGC12},
control-plan reputation \cite{Konte:2015:ARS:2785956.2787494}, early-detection time \cite{Konte:2015:ARS:2785956.2787494}, cybersecurity posture \cite{XuInTrust2014}, sweep-time \cite{XuInTrust2014}, attack rate \cite{XuTIFS2013,XuTIFS2015} & comprehensive cyber threat posture$^\dag$ \\
~~zero-day attacks & number of zero-day attacks \cite{Symantec-0-day-report-April-2012}, lifetime of zero-day attacks  \cite{Bilge:2012:BWK:2382196.2382284}, number of zero-day attack victims \cite{Bilge:2012:BWK:2382196.2382284} & susceptibility of a computer to zero-day attacks$^\dag$ \\
~~attack power && \\
~~~~power of targeted attacks & targeted threat index  \cite{hardy2014targeted} & susceptibility to targeted attacks$^\dag$\\
~~~~power of botnet & botnet size \cite{DagonNDSS06}, botnet efficiency \cite{DBLP:conf/acsac/DagonGLL07}, botnet robustness \cite{DBLP:conf/acsac/DagonGLL07} & botnet attack power$^\dag$, botnet resilience with counter-countermeasures$^\dag$ \\
~~~~power of malware spreading & infection rate \cite{DBLP:conf/infocom/ChenJ07} & attack power$^\dag$, wasted scans$^\dag$ \\
~~~~power of multi-stage attacks &
necessary defense \cite{Sheyner:2002:AGA:829514.830526}, weakest adversary \cite{PamulaQoP2006}, attack paths \cite{Ritchey:2000:UMC:882494.884423,Sheyner:2002:AGA:829514.830526,Jha:2002:TFA:794201.795177,OuCyberSA2014}, $k$-zero-day-safety \cite{WangESORICS08}, effort-to-security-failure \cite{DacierIFIPSEC2006,OrtaloIEEETSE2009} & multi-stage attack power$^\dag$ \\
~~~~power of evading detection & (no nontrivial metrics defined) & evasion capability$^\dag$ \\~~obfuscation sophistication & obfuscation prevalence  \cite{Roundy:2013:BOP:2522968.2522972}, packer structural complexity \cite{DBLP:conf/sp/Ugarte-PedreroB15} & obfuscation sophistication$^\dag$ \\
 \hline\hline
\multicolumn{3}{|l|}{Measuring situations} \\ \hline
~~security state & fractions of compromised computers at time $t$, probability a computer is compromised at time $t$ \cite{SandersQUEST2011,DBLP:conf/hotsos/DaXX14,XuCybersecurityDynamicsHotSoS2014} & $S(t)$ and $s_i(t)$ for any $t$ $^\dag$ \\
~~security incidents & &\\
~~~~incident spatial characteristics &  incident rate \cite{CIS,MicrosoftIntelligenceReport,Maier:2011:AOM:2026647.2026660,DBLP:conf/ccs/YenHORJ14,LalondeLevesque:2013:CSR:2508859.2516747}, encounter rate \cite{DBLP:conf/ccs/YenHORJ14,Mezzour:2015:ESG:2746194.2746202,MicrosoftIntelligenceReport,Maier:2011:AOM:2026647.2026660,LalondeLevesque:2013:CSR:2508859.2516747}  & incident occurrence frequency$^\dag$ \\
~~~~incident temporal characteristics & delay in incident detection \cite{CIS}, time between incidents \cite{CIS,JonssonIEEETSE1997,TrivediDSN2002,HolmIEEETDSC2014}, time-to-first-compromise \cite{JonssonIEEETSE1997,TrivediDSN2002,HolmIEEETDSC2014} & predictive incident occurrence frequency$^\dag$ \\
~~~~incident damage & cost of incidents \cite{CIS} & predictive incident damage$\dag$  \\
~~security investments &  security spending \cite{NIST800-55Rev1,CIS}, security budget \cite{CIS} & payoff of security investment$^\dag$\\
\hline
\end{tabular}
}
\end{minipage}
}
\label{table:summary-of-metrics}
\end{sidewaystable}

\subsection{What should we measure?}


First, Table \ref{table:summary-of-metrics} shows that there are big gaps between the state-of-the-art metrics (i.e., second column) and the desirable metrics (i.e., third column).
For example,  we used the thickness of blue bars and red arrows in Figures \ref{fig:taxonomy} and \ref{fig:taxonomy-individual-view}
to reflect the defense power and attack power, respectively. However, the existing security metrics do not measure this intuitive thickness metric.
This resonates Pfleeger's observation \cite{Pfleeger:2009:UCM:1591880.1592054} that
metrics in the literature often correspond to ``what {\em can be} easily measured,'' rather than ``what {\em need to be} measured''---a fundamental problem that is largely open.
In what follows we discuss the 4 categories systemized above and highlight what kinds of research are needed.

For measuring system vulnerabilities, we considered metrics for measuring users' vulnerabilities, password vulnerabilities,
system interface-induced vulnerabilities, software vulnerabilities, and cryptographic key vulnerabilities.
These classes of metrics appear to be complete. For example, the problem of insider threats could be treated by some users' vulnerability metrics,
such as {\em user's susceptibility to insider threats}. (The survey did not include security metrics for measuring insider threats,
simply because there are no well-defined metrics of this kind despite the efforts \cite{7010900,Martinez-Moyano:2008:BTI:1346325.1346328}.)
However, it is not clear what kind of formalism would be sufficient for reasoning about the completeness of metrics.
A related open problem is: How can we define a metric that may be called {\em overall vulnerability} of an enterprise or computer system,
which reflects the systems' overall susceptibility to attacks?

For measuring defense power, we considered metrics for measuring the effectiveness of blacklisting,
the attack detection power, the effectiveness of ASLR, the effectiveness of assuring CFI, and the overall defense power.
It is ideal that the overall defense power metric can accommodate the other kinds of metrics.
An important open problem is: Can the study of the overall defense power metric help determine the completeness of the
other classes of defense power metrics? For example, is there a formalism by which we can rigorously show that the overall
defense power metric can or cannot be derived from the other kinds of metrics?

For measuring threats, we considered metrics for measuring threat landscape, zero-day attacks, attack power, and obfuscation sophistication.
The question is: How can we rigorously show that these metrics collectively reflect the intuitive metric that may be called {\em overall attack power}?

For measuring situations, we considered metrics for measuring the global security state $S(t)=(s_1(t),\ldots,s_n(t))$, security incidents, and security investments.
These metrics appear to be complete because security state $S(t)$ itself does not reflect the damage of security incidents, which may or may not be positively
correlated to security investments. Nevertheless, it may be helpful to integrate these metrics and the metrics for measuring system vulnerabilities and defense power
into a single category, which may more comprehensively reflect the situations. This is because, for example, a user's susceptibility to attacks may vary with time $t$.
An important open problem is: How can the defense power metrics and the attack power metrics be unified into a single framework
such that, for example, a single algorithm would allow us to compute the outcome (or consequence) of the interaction
between an attacker of a certain attack power and a defender of a certain defense power?
This would formalize the intuitive representation of attack power and defense power as highlighted in Table \ref{table:summary-of-metrics} (third column),
namely the thickness of blue bars and red arrows in Figures \ref{fig:taxonomy} and \ref{fig:taxonomy-individual-view}.
Resolving this problem would immediately lead to a formal treatment of the arms race between the attacker and the defender,
as exemplified by the discussion on the effectiveness of ASLR (Section \ref{sec:defense-power-ASLR}) and CFI (Section \label{sec:defense-power-CFI}).

Second, what would be the {\em complete} set of security metrics from which any useful security metric can be derived?
The concept of {\em completeness} not only applies to the categories of security metrics, but also applies to the security metrics within each category.
In order to shed light on this fundamental problem, let us look at the example of healthcare.
In order to determine a person's health state, various kinds of blood tests are conducted to measure things such as glucose.
The tests are subsequential to the medical research that discovered (for example) that glucose is reflective of a certain aspect of
human body's health state, which answers the {\em what to measure} question.
This example highlights that more research is needed to understand {\em what security metric is reflective of which security attribute or property};
otherwise, our understanding of security metrics will remain heuristic.

\subsection{How to measure what we should measure in practice: random variables or numbers?}

Security metrics are difficult to measure in practice.
For example, vulnerabilities are dynamically discovered and
the attacker may identify some zero-day vulnerabilities that are not known to the defender;
the defender does not know for certain what exploits the attack possess;
there may be some attack incidents that are never detected by the defender.
These indicate that uncertainty is inherent to the threat model the defender is confronted with.
As a consequence, security metrics should often be treated as random variables, rather than numbers.
This means that we should strive to characterize the distributions of the random variables representing security metrics, rather than the
means of random variables only. Another uncertainty is caused by the measurement error,
such as $S(t)\neq O(t)$ as illustrated in Figure \ref{fig:taxonomy}.
This further highlights the importance of treating security metrics via random variables.

\subsection{To aggregate, or not to aggregate?}

The need to consider security at multiple levels/aspects of abstractions can be at least traced back to Lampson \cite{Lampson06practicalprinciples}
in stopping control-flow hijack attacks \cite{DBLP:conf/uss/CarliniBPWG15}.
It is ideal to aggregate multiple security metrics into a single one.
But, the problem is {\em how} \cite{5432146,Pfleeger:2009:UCM:1591880.1592054} and {\em to what extent}.
For the {\em how} part of the problem, the difficulty lies in the need to tackle the {\em dependence} between the security metrics,
which often exists because the same system with the same vulnerabilities and the same defense deployment would get compromised by the same attack.
For example, there may exist some dependence between the {\em security investment}, {\em security coverage}, and {\em damage of incidents}.
When the mean of random variables (representing metrics) is the only concern,
we can indeed aggregate multiple measures into a single one via some linear combination of them.
However, the mean of random variables is just one decision-making factor, because we often need to consider
the variance of the random variable in question.
This requires us to tackle the dependence between the measures.
For the {\em to what extent} part of the problem, the difficulty lies in the determination of degree of aggregation.
At the highest level of aggregation, one may suggest to have metrics such as {\em degree of confidentiality} and
{\em degree of integrity}. Even if it is feasible to do so, the operational utilities of such metrics would be limited
because they typically measure the outcome of attack-defense interactions and do not necessarily give guidance
for adjusting the defense during the course of attack-defense interactions.

\subsection{What are the desirable properties of security metrics?}

There have been proposals for characterizing ``good'' metrics, such as the following.
From a conceptual perspective, a good metric should be {\em easy to understand} not only to researchers, but also to defense operators \cite{MIT-Lincoln-Lab-Metrics-TR-2012}.
From a measurement perspective, a good metric should be {\em relatively easy to measure}, consistently and repeatably  \cite{Jaquith:2007:SMR:1214710}.
From a utility perspective, a good metric should allow both {\em horizontal comparison} between enterprise systems,
and {\em temporal comparison} (e.g., an enterprise system in the present year vs the same enterprise system in the last year) \cite{DepartmentofState2010,MIT-Lincoln-Lab-Metrics-TR-2012}.

However, we also need to understand the mathematical properties security metrics should possess.
These properties not only can help us differentiate the good metrics from the bad metrics,
because for example we can conduct transformations between metrics.
These properties may also ease the measurement process.
To see this, let us look at the particular property of {\em additivity}.
The property of {\em additivity} can be understood from the following example.
When we talk about the measurement of mass, we are actually seeking a mapping $mass$ from $\A=(Objects,\text{\sf
heavier-than},\op)$ to $\B=(\R^+\cup\{0\},>,+)$, where $\op$ can be the
``putting together" operation and $\R^+$ is the set of positive
reals. Then, {\em mass} should satisfy: (i) For two objects $a$ and
$b$, if $a\ \text{\sf heavier-than}\ b$, then $mass(a)>mass(b)$.
(ii) For any objects $a$ and $b$, $mass(a \op b)=mass(a)+mass(b)$.
Although the above (i) is relatively easy to achieve when measuring security,
the above (ii), namely the additivity property, rarely holds in this domain.
However, the additivity is useful because it substantially eases the measurement operations.
A partial explanation for the lack of additivity is that security exhibits emergent behavior \cite{5432146,XuEmergentBehaviorHotSoS2014}.
This suggests to investigate whether there are some additivity-like properties that can help ease the measurement of security.

\subsection{What partnership is needed to tackle the problem?}

The problem of security metrics may not be solved without a good partnership between the government, industry, and academia.
We strongly suggest that whenever possible, any security publication in the future should include an {\bf explicit} definition of the security metric that is
relevant to the study. The reality is that most security publications did not define the security metrics they use.
For example, a defense mechanism is often published with a security analysis that often makes a qualitative statement like
``the mechanism can defeat a specific attack in a specific threat model.''
Because of the diversity in terms of the kinds of security metrics, both {\em bottom-up} and {\em top-down} approaches are important.
For the bottom-up approach, each publication should strive to define, as precise as possible, the security metric
and its attributes. In terms of attributes of security metrics, one can consider
its {\em temporal} characteristics, its {\em spatial} characteristics, and its connection to high-level security concepts such as
confidentiality, integrity and availability.
Moreover, we strongly suggest that the security curriculum should have substantial materials for educating
and training future generations of security researchers and practitioners with a systematic body of knowledge in security metrics.
This is largely hindered by the lack of systematic treatment.
We hope the present survey can catalyst the development of such materials, which however might not be possible until after
security publications include serious effort at explicitly defining the security metrics in question.

While academic researchers are perhaps obligated to propose {\em what to measure}, they
often do not have access to real data.
The industry has the data,
but is often prohibited from sharing data with academic researchers because of legitimate concerns (e.g., privacy).
Although the government has already incentivized data sharing through projects such as {\tt PREDICT} (www.predict.org),
our research experience hints that semantically richer data is imperative for tackling the problem of security metrics.